\documentclass[journal]{IEEEtran}
\usepackage[latin9]{inputenc}
\usepackage{amsmath}
\usepackage{amssymb}
\usepackage{graphicx}
\usepackage{esint}

\makeatletter

\providecommand{\tabularnewline}{\\}

\ifCLASSINFOpdf
\else
\fi
\usepackage{hyperref}
\hypersetup{
    colorlinks=true,
    linkcolor=blue,
    filecolor=magenta,      
    urlcolor=cyan,
}

\makeatother

\begin{document}

\title{Multi-Slot Over-The-Air Computation in Fading Channels }
\author{\IEEEauthorblockN{Suhua TANG$^{1}$, Petar Popovski$^{2}$, Chao
ZHANG$^{3}$, Sadao OBANA$^{1}$} \\
\IEEEauthorblockA{ $^{1}$Graduate School of Informatics and Engineering,
The University of Electro-Communications, Japan\\
 $^{2}$Department of Electronic Systems, Aalborg University, Denmark\\
$^{3}$School of Information and Communications Engineering, Xi\textquoteright an
Jiaotong University, China\\
Email: shtang@uec.ac.jp}}
\maketitle
\begin{abstract}
IoT systems typically involve separate data collection and processing,
and the former faces the scalability issue when the number of nodes
increases. For some tasks, only the result of data fusion is needed.
Then, the whole process can be realized in an efficient way, integrating
the data collection and fusion in one step by  over-the-air computation
(AirComp). Its shortcoming, however, is signal distortion when channel
gains of nodes are different, which cannot be well solved by transmission
power control alone in times of deep fading. To address this issue,
in this paper, we propose a multi-slot over-the-air computation (MS-AirComp)
framework for the sum estimation in fading channels. Compared with
conventional data collection (one slot for each node) and AirComp
(one slot for all nodes), MS-AirComp is an alternative policy that
lies between them, exploiting multiple slots to improve channel gains
so as to facilitate power control. Specifically, the transmissions
are distributed over multiple slots and a threshold of channel gain
is set for distributed transmission scheduling. Each node transmits
its signal only once, in the slot when its channel gain first gets
above the threshold, or in the last slot when its channel gain remains
below the threshold. Theoretical analysis gives the closed-form of
the computation error in fading channels, based on which the optimal
parameters are found. Noticing that computation error tends to be
reduced at the cost of more transmission power, a method is suggested
to control the increase of transmission power. Simulations confirm
that the proposed method can effectively reduce computation error,
compared with state-of-the-art methods.
\end{abstract}


\section{Introduction}


Internet of Things (IoT) will have billions of devices, many of which
will be connected to the Internet by the low power wide area (LPWA)
technologies such as NB-IoT and LoRa \cite{Sinha17}. Data collection
and processing in IoT systems usually are separated, both in the digital
domain, and the former involves the one-by-one data transmission.
Then, a network with $K$ nodes will involve at least $K$ transmissions
per data collection, which takes much time when there are millions
of nodes in a LPWA cell. This is a challenging problem, because of
limited spectrum bandwidth, real-time requirement, and the transmission
collisions in a multiple access channel.

Some sensing tasks actually do not require the individual value from
each sensor, if only their fusion, e.g., sum, average, max, etc.,
is computed correctly. For these tasks, a more efficient method is
possible. Recently, a new policy, called over-the-air computation
(AirComp) \cite{Nazer07}, was investigated, which usually integrates
the data collection and fusion in the analog domain using uncoded
transmissions \cite{Xiao08}. All nodes transmit their signals simultaneously
in a coordinated way and the data fusion (sum) is computed over the
air. In addition, it has been proven that the computation error of
the analog transmission can be made much smaller than that of digital
methods when using the same amount of resources \cite{CZLee20}.

Although AirComp generally only supports the sum operation, by proper
pre-processing and post-processing operations, it can be extended
to support any kind of nomographic functions \cite{Buck79,Goldenbaum13c,Abari16}.
Recently, deep neural network is also exploited in the pre-processing
and post-processing, which enables more advanced processing of sensor
data by approximating any function via deep models learned from data
\cite{HaoYe20}.

AirComp only takes one slot for data collection, which is very efficient.
But to ensure the unbiased data fusion, each node has to pre-amplify
its data so that signals arriving at the sink are aligned in their
signal magnitude \cite{Liu20,Cao19}. The pre-amplification usually
uses a principle of channel inversion (a power inversely proportional
to the channel coefficient) to mitigate the difference in channel
gains. This causes much transmission power in deep fading. Few works
have studied this problem, especially in the fast fading environment. 

To solve this problem, in this paper, we propose a multi-slot over
the air computation (MS-AirComp) framework for the sum estimation
in fading channels. Conventional data collection and AirComp are two
extremes. The former uses one slot for each node and the latter uses
one slot for all nodes. MS-AirComp is an alternative policy that lies
between them, exploiting multiple slots to take advantage of time-varying
channel gains so as to facilitate power control. It is necessary to
avoid transmission when a node is in deep fading. This is also feasible
when considering the time diversity due to the random variation of
channel gain \cite{Goldsmith05}. Then, the over-the-air computation
is distributed over $N$ slots, where $N$ is much less than that
of nodes. Specifically, the sink node sets a threshold of channel
gain, based on the channel statistics. Each node transmits its signal
only once in the $N$ slots, either in the slot when  its channel
gain first gets above the predefined threshold, or  in the last slot
if its channel gain remains below the threshold. The transmission
power control policy is the same as in previous methods \cite{Liu20,Cao19},
either the channel inversion policy if the power is less than the
constraint or using the maximal value otherwise. 

The main contributions of this paper are 
\begin{itemize}
\item Shaping channel gain by exploiting time diversity. This paper improves
channel gains by using multiple slots and setting a threshold of channel
gain. This helps to avoid deep fading and facilitate power control
to achieve signal magnitude alignment.
\item Distributed transmission scheduling. Transmission scheduling of each
node is performed in a distributed way, using a threshold of channel
gain, which the sink computes based on the channel statistics instead
of instantaneous values.
\item Closed-form of the computation mean squared error (MSE). Theoretical
analysis of the computation MSE enables to find optimal parameters
given the statistics of channel gains, and the simulation results,
based on these optimal parameters, are consistent with the analysis.
 
\item Revealing the tradeoff between the computation MSE and transmission
power and presenting a simple solution for this issue.
\end{itemize}
Numerical analysis and Monte Carlo simulations illustrate the promising
performance of the proposed method. Compared with state-of-the-art
methods, the proposed method greatly reduces the computation MSE,
by improving channel gain at the cost of only a few slots. By a refined
tradeoff between the computation MSE and transmission power, the proposed
method effectively reduces the computation MSE while consuming almost
the same transmission power as the previous method. 

The rest of this paper is organized as follows: Sec.~\ref{sec:Related}
reviews the basic AirComp method and related work. Sec.~\ref{sec:Multislot-SingleTx}
first presents the proposed framework, and analyzes the computation
MSE. On this basis, optimal parameters are found, transmission power
is analyzed, and some numerical results are illustrated. Sec.~\ref{sec:Evaluation}
shows the results achieved by Monte Carlo simulation, and points out
the necessity of a tradeoff between the computation MSE and transmission
power. Then, in Sec.~\ref{sec:tradeoff-pow-mse}, a simple method
is suggested to control the increase of transmission power. Finally,
Sec.~\ref{sec:Conclusion} concludes this paper and points out future
work.

\section{Related Work\label{sec:Related}}

Here, we review the basic over-the-air computation method, and previous
efforts on improving its performance. 

\begin{table}[!t]
\caption{Main notations for analysis.}

\label{tab:SimCond} 

\begin{centering}
\begin{tabular}{l|l|l}
\hline 
Notation & Meaning & Default value\tabularnewline
\hline 
$K$ & Number of sensor nodes & 100\tabularnewline
$h_{k}$ & Channel coefficient of node $k$ & \tabularnewline
$g_{k}$ & Channel gain of node $k$  & \tabularnewline
$\overline{g}_{k}$ & Average channel gain of node $k$ & 10\tabularnewline
$b_{k}$ & Tx-scaling factor for node $k$  & \tabularnewline
$P_{max}$ & Normalized maximal Tx power & 10\tabularnewline
$g_{th}$ & Threshold of channel gain & \tabularnewline
$P_{th}$ & Threshold of signal receive rate & \tabularnewline
\hline 
\end{tabular}
\par\end{centering}
\end{table}

\subsection{Basic AirComp model\label{sec:Basic-AirComp}}

AirComp was originally studied for sensor networks.  Here we review
a typical AirComp model \cite{Liu20}. In a sensor network, $K$ nodes
each transmit the analog signals to a common sink, simultaneously
in the same slot, triggered by a beacon signal from the sink. The
sink tries to compute the sum of signals received from all nodes.
Both the nodes and the sink have a single antenna. To deal with the
difference in channel gains, the pre-processed signal at the $k_{th}$
node, $x_{k}\in\mathbb{C}$, with zero mean and unit variance ($E(|x_{k}^{2}|)=1$),
is amplified by its Tx-scaling factor $b_{k}\in\mathbb{C}$ and sent
to the sink. All the transmissions are synchronized so that all signals
arrive at the sink at the same time. The sink applies a Rx-scaling
factor $a\in\mathbb{C}$ to the received signal to get the computation
result as

\begin{equation}
r=a\cdot\left(\sum_{k=1}^{K}h_{k}b_{k}x_{k}+n\right),\label{eq:1slot-model}
\end{equation}
where $h_{k}\in\mathbb{C}$ is the channel coefficient between node
$k$ and the sink, and $n\in\mathbb{C}$ is the additive white Gaussian
noise (AWGN) at the sink with zero mean and variance being $\sigma^{2}$.
Channel coefficient $h_{k}$ is assumed to be known by node $k$ and
the sink (assuming that there is a pilot signal for channel measurement
and that the channel is semi-static). 

With the maximal power constraint, $|b_{k}x_{k}|^{2}$ should be no
more than $P'$, the maximal power. Let $P_{max}$ denote $P'/v^{2}$.
Then, $b_{k}^{2}\le P'/v^{2}=P_{max}$. In the case of unbiased estimation
of the signal sum, $h_{k}b_{k}$ should have a constant value $\alpha$,
so that $\sum_{k}x_{k}$ can be estimated from Eq.(\ref{eq:1slot-model}).
This is called the channel inversion policy \cite{Liu20,Cao19} because
$b_{k}$ is computed as $\alpha/h_{k}$ with a parameter $\alpha$. 

Parameters $b_{k}$ and $a$ are set to make $r$ approach the target
sum $\sum_{k=1}^{K}x_{k}$. This is achieved by minimizing the following
computation MSE (under the power constraint $|b_{k}|^{2}\le P_{max},k=1,\cdots,K$),
where $E\{\cdot\}$ is the expectation operation.

\begin{equation}
MSE\negmedspace=\negmedspace E\left\{ |r\negmedspace-\negmedspace\sum_{k=1}^{K}x_{k}|^{2}\right\} \negmedspace=\negmedspace\sum_{k=1}^{K}|ah_{k}b_{k}\negmedspace-\negmedspace1|^{2}\negmedspace+\negmedspace\sigma^{2}|a|^{2}.\label{eq:mse-aircomp}
\end{equation}
Node $k$ can always adjust $b_{k}$ to ensure that $h_{k}b_{k}$
is real and positive. Therefore, in the following, it is assumed that
$h_{k}\in\mathbb{R^{+}}$, $b_{k}\in\mathbb{R^{+}}$, and $a\in\mathbb{R^{+}}$,
for the simplicity of analysis.

\subsection{Dealing with signal distortion in AirComp}

Several factors may affect the performance of AirComp.  Timing synchronization
usually is a necessity of AirComp for avoiding signal distortion.
By modulating the sensor data in a series of random signal pulses,
AirComp can be realized by a coarse block-synchronization \cite{Goldenbaum13a}.
In \cite{Abari15}, AirShare is proposed, in which the sink synchronizes
the clock of all nodes before the actual transmission, by using multiple
frequencies. 

Signal distortions also may be caused by noise and channel fading.
Distortion outage for AirComp was investigated in \cite{Wang11},
where an outage is defined as an event in which the estimation error
exceeds a predefined threshold. In deep fading, the magnitudes of
some signals cannot be aligned with that of other signals, under the
constraint of maximal transmission power. Some efforts have been devoted
to solving this problem. The work in \cite{Liu20} studies the power
control policy, aiming to minimize the computation error by jointly
optimizing the transmission power and a receive scaling factor at
the sink node. Specifically, for a node $k$, its power is computed
as $b_{k}=\alpha/h_{k}$ if $h_{k}$ is large enough so that $b_{k}$
is below the power constraint. Otherwise, the maximal power is used.
In \cite{Cao19}, the authors further consider the time-varying channel
by regularized channel inversion, aiming at a better tradeoff between
the signal-magnitude alignment and noise suppression. Antenna array
was also investigated in \cite{Zhu19H,Wen19} to support vector-valued
AirComp.  It is further combined with wireless power transfer in
\cite{Li19ZGH}. 

Recently, AirComp is also applied to model aggregation in federated
learning, where a global model, shared among nodes, is locally updated.
The model updates are then aggregated to form a new model. When all
nodes are connected to a server by a wireless channel, the model collection
and aggregation can be integrated by AirComp.

To deal with fading and noise in wireless channel, the authors in
\cite{Sery20} investigated direct model update based on the noisy
distorted gradient. Precoding and scaling operations are suggested
to mitigate the effect of the noisy channel to accelerate the convergence
of the learning process \cite{TomerSery20}. In federated learning,
it is not necessary to receive model update from all nodes. Therefore,
in \cite{Amiri20}, at each iteration, only nodes with a channel gain
large enough are selected to transmit their model update.  Fast fading
is taken into account in \cite{Frey2020}, for the purpose of one-shot
approximation of function values, considering sub-Gaussian noise and
correlated channels.

\subsection{Dealing with fading in multiple access channel}

In the multiple access channel, it is efficient to exploit multi-user
diversity to deal with channel fading, letting nodes with high channel
gains transmit their signals first. This is studied for ALOHA networks
in \cite{Qin06,Adireddy05}, and for CSMA networks in \cite{tang14}.
Generally, a threshold of channel gain is set for nodes to facilitate
distributed transmission scheduling. But these methods only let one
node transmit its signal each time in order to avoid transmission
collisions.

\subsection{A short comparison}

Fast channel fading is a large challenge for AirComp, but has not
been well studied yet. Considering the specific task of model aggregation
in federated learning, the policy of node selection is adopted in
\cite{Amiri20}, which relies on the centralized control, both to
obtain channel gains and to get the correct number of nodes involved
in the transmission. But for a general sensor network, data from all
nodes need to be collected. In addition, it is impractical for the
sink to collect channel gain of all nodes in some cases. Transmission
power control \cite{Liu20,Cao19} is necessary in AirComp, but its
effect is limited in times of deep fading. In comparison, the proposed
method improves the channel gain by leveraging the time diversity,
letting nodes schedule their transmissions only when their channel
gains are high enough. This distributes over-the-air computation into
multiple slots, improving system performance at the cost of only a
few slots. 

\section{Multi-Slot AirComp \label{sec:Multislot-SingleTx}}

Here, we present the proposed MS-AirComp framework, and analyze the
computation MSE using the distribution of channel gains. Then, we
further analyze transmission power etc., and show some numerical results.

\subsection{System model\label{sec:Framework}}

This paper focuses on the sum estimation of sensing data obtained
by all nodes in the sensor network. It can be easily extended to support
other computations by proper pre-processing and post-processing. The
unbiased, low-noise estimation is desired. To this end, the transmission
is distributed to multiple slots.

\begin{figure}
\centering

\includegraphics[width=8cm]{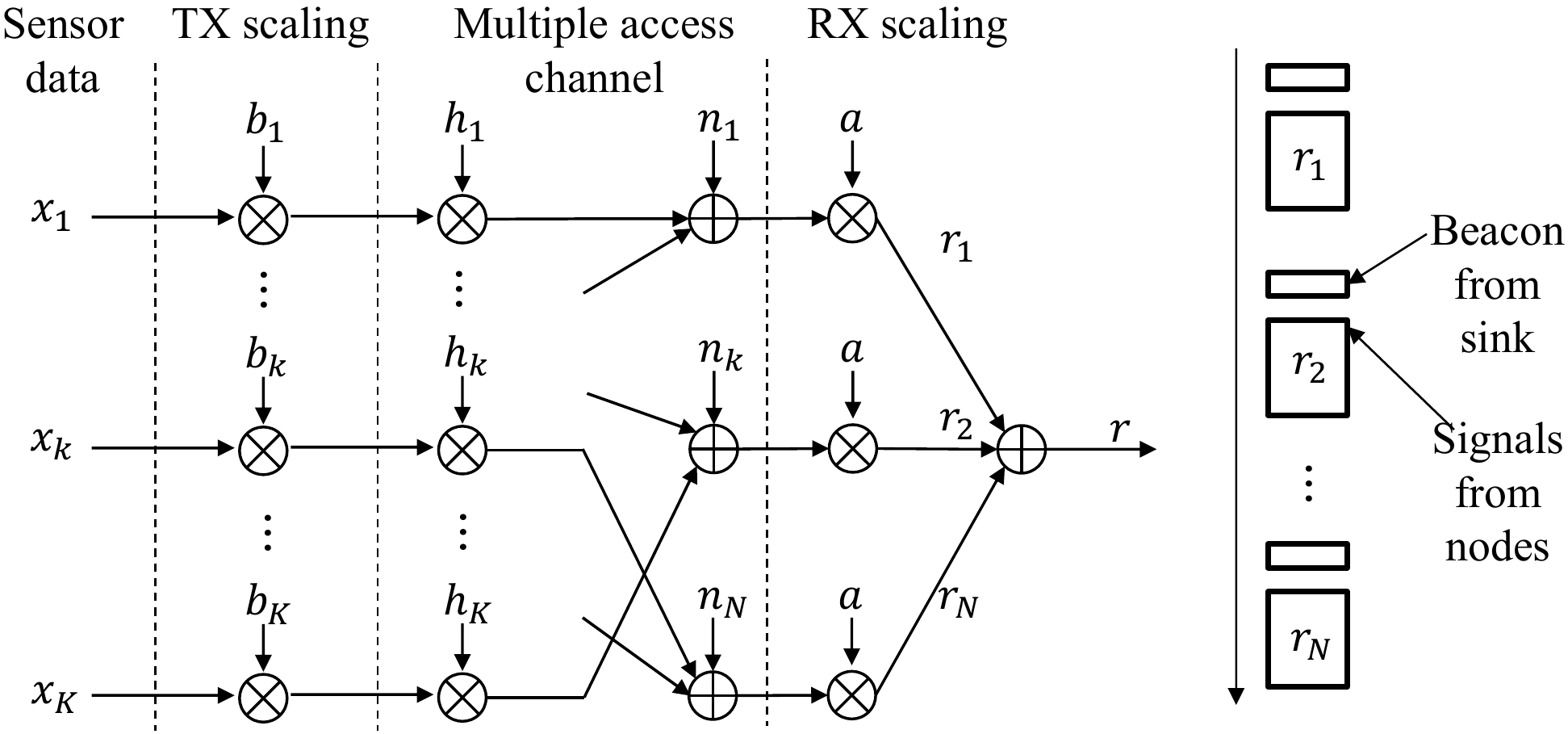}

\caption{\label{fig:MultiSlotModel}Proposed system model with $K$ nodes and
1 sink. All nodes should transmit their signals to the common sink
within $N$ slots, but each node may select which slot to transmit
based on its channel gain. }
\end{figure}

Figure~\ref{fig:MultiSlotModel} shows the system model. We investigate
a multi-slot over-the-air computation (MS-AirComp) framework. Specifically,
there are $N$ slots for the whole transmission, and each node has
the chance to choose a slot with a high channel gain to transmit its
signal, $x_{k}\in\mathbb{C}$, with zero mean and unit variance ($E(|x_{k}^{2}|)=1$).
At the beginning of each slot, the sink broadcasts a beacon signal,
based on which each node detects channel gain and synchronizes its
clock with the sink. 

We assume the channel experiences fast fading and channel gain independently
varies per slot. Different from the model in Sec.~\ref{sec:Basic-AirComp},
instantaneous channel coefficient $h_{k}\in\mathbb{C}$ is known by
node $k$ (by receiving the beacon signal) but not by the sink (assuming
a dynamic channel without immediate feedback of channel coefficient
from node $k$ to the sink). Instead, the sink learns the channel
statistics, and on this basis decides the optimal parameters.  The
same Rx-scaling factor $a\in\mathbb{C}$ is used for all slots, assuming
the channel is stationary. Each node can monitor $N$ channel coefficients
in $N$ slots. But for the simplicity of analysis, here $h_{k}$ is
used to represent the channel gain of the slot over which node $k$
actually transmits its signal, and $b_{k}\in\mathbb{C}$ corresponds
to the transmission power. 

The signals received over $N$ slots, summed together at the sink,
are 

\begin{equation}
r=a\cdot\left(\sum_{k=1}^{K}h_{k}b_{k}x_{k}+\sum_{i=1}^{N}n_{i}\right).\label{eq:Nslot-model}
\end{equation}
The signal sum has the same form as in the single-slot case in Eq.(\ref{eq:1slot-model}),
except that the noises in all slots are involved. It is expected that
the improvement in channel gain is greater than the increase in noise
power. Similar to the AirComp method, it is assumed that $h_{k}\in\mathbb{R^{+}}$,
$b_{k}\in\mathbb{R^{+}}$, and $a\in\mathbb{R^{+}}$ for the simplicity
of analysis. 

The computation MSE is defined in a similar way as in Eq.(\ref{eq:mse-aircomp}),
as follows 
\begin{eqnarray}
MSE & \negmedspace\negmedspace=\negmedspace\negmedspace & E\left\{ \sum_{k=1}^{K}\left(a\cdot\alpha_{k}-1\right)^{2}\right\} +Na{}^{2}\sigma^{2},\label{eq:Nslot-mse}\\
\alpha_{k} & \negmedspace\negmedspace=\negmedspace\negmedspace & h_{k}b_{k}.\nonumber 
\end{eqnarray}
But here $\alpha_{k}$ itself is random, so the expectation operator
is kept. 

Assume on each slot, $\alpha_{k}$ has a probability density function
(PDF) $f_{\alpha_{k}}(x)$ and a cumulative distribution function
(CDF) $F_{\alpha_{k}}(x)$, and their counterparts over $N$ slots,
under a selection policy, are $f_{\alpha_{k}}^{N}(x)$ and $F_{\alpha_{k}}^{N}(x)$,
respectively. Let $\delta(x)$ represent the impact of signal magnitude
alignment, and $\delta(\alpha_{k})$ denote the actual value of $\alpha_{k}$.
Then, the computation MSE in Eq.(\ref{eq:Nslot-mse}) is computed
as the expectation with respect to $\alpha_{k}$, as follows:

\begin{equation}
MSE\negmedspace=\negmedspace\sum_{k=1}^{K}\intop_{0}^{\infty}\left(a\cdot\delta(x)\negmedspace-\negmedspace1\right)^{2}f_{\alpha_{k}}^{N}(x)dx\negmedspace+\negmedspace Na^{2}\sigma^{2}.\label{eq:mse-probmodel}
\end{equation}

\subsection{Impact of signal magnitude alignment}

In the transmission, magnitudes ($\alpha_{k}$) of signals from most
nodes will be aligned to a common value, which is denoted as $\alpha_{th}$.
The Tx-scaling $b_{k}$ used for the transmission at node $k$ is
$b_{k}=min(\alpha_{th}/h_{k},\sqrt{P_{max}})$, $\alpha_{th}/h_{k}$
if it is no more than $\sqrt{P_{max}}$ (using the channel inversion
policy) and $\sqrt{P_{max}}$ otherwise (using the maximal power).
This power control policy is the same as in the previous work \cite{Liu20}.
The difference is that $h_{k}$ in the actual transmission, after
slot selection, is improved in the proposed method.

Considering the above policy of power control, in the following we
consider $\alpha_{k}=\sqrt{P_{max}}h_{k}$, the maximal available
signal magnitude. With channel gain $g_{k}$, $h_{k}=\sqrt{g_{k}}$,
then $\alpha_{k}=\sqrt{P_{max}g_{k}}$. 

At current channel gain $g_{k}$, node $k$ needs to check whether
$\sqrt{P_{max}g_{k}}$, at the maximal transmission power, reaches
the threshold $\alpha_{th}$. If affirmative, its signal magnitude
will be aligned to $\alpha_{th}$, possibly using a smaller transmission
power.  Then, $\delta(\alpha_{k})$, representing the value of $\alpha_{k}$
after signal magnitude alignment, corresponds to the function

\begin{equation}
\delta(x)=\begin{cases}
x & 0\le x<\alpha_{th},\\
\alpha_{th} & x\ge\alpha_{th}.
\end{cases}
\end{equation}
This applies to the selected slot where a node transmits its own signal.
Using this function, $\left(a\cdot\delta(x)\negmedspace-\negmedspace1\right)^{2}$
in Eq.(\ref{eq:mse-probmodel}) becomes $\left(a\cdot x\negmedspace-\negmedspace1\right)^{2}$
in the range $(0,\alpha_{th})$, and $\left(a\cdot\alpha_{th}\negmedspace-\negmedspace1\right)^{2}$
in the range $(\alpha_{th},\infty)$ with a probability $1\negmedspace-\negmedspace F_{\alpha_{k}}^{N}(\alpha_{th})$.
Then, the computation MSE in Eq.(\ref{eq:mse-probmodel}) is simplified
as

\begin{align}
MSE\negmedspace & =\negmedspace\sum_{k=1}^{K}MSE(k)\negmedspace+\negmedspace Na^{2}\sigma^{2},\label{eq:mse-optsel}\\
MSE(k)\negmedspace=\negmedspace\negmedspace & \intop_{0}^{\alpha_{th}}\negmedspace\negmedspace\negmedspace\left(a\cdot x\negmedspace-\negmedspace1\right)^{2}\negmedspace f_{\alpha_{k}}^{N}(x)dx\negmedspace+\negmedspace\left(a\cdot\alpha_{th}\negmedspace-\negmedspace1\right)^{2}\negmedspace\left(1\negmedspace-\negmedspace F_{\alpha_{k}}^{N}(\alpha_{th})\right).\nonumber 
\end{align}
The target is to find the optimal parameters $a$ and $b_{k}$ that
minimize the computaion MSE, under the power constraint.

\begin{align}
\underset{a,b_{k}}{min} & \ MSE\\
s.t. & \ b_{k}^{2}\le P_{max},k=1,\cdots,K.\nonumber 
\end{align}
The computation MSE is a quadratic function of $a$. By letting the
partial differentiation $\frac{\partial MSE}{\partial a}$ equal to
0, the optimal $a$ can be computed via Eq.(\ref{eq:opt_a_optsel}).

\begin{figure*}
\begin{equation}
a_{opt}=\frac{\sum_{k=1}^{K}\left(\intop_{0}^{\alpha_{th}}xf_{\alpha_{k}}^{N}(x)dx+\alpha_{th}\left(1-F_{\alpha_{k}}^{N}(\alpha_{th})\right)\right)}{\sum_{k=1}^{K}\left(\intop_{0}^{\alpha_{th}}x^{2}f_{\alpha_{k}}^{N}(x)dx+\alpha_{th}^{2}\left(1-F_{\alpha_{k}}^{N}(\alpha_{th})\right)\right)+N\sigma^{2}}.\label{eq:opt_a_optsel}
\end{equation}
\end{figure*}

It is possible to further compute the partial differentiation $\frac{\partial MSE}{\partial\alpha_{th}}=0$,
but it has no closed-form for $\alpha_{th}$. Therefore, $\alpha_{th}$
is found by minimizing the computation MSE via grid search.

\subsection{Selecting a slot by a threshold}

In the proposed method, we use a thresholding policy for distributed
transmission scheduling, and a threshold of channel gain $g_{th}$
is set. In each slot, based on the beacon signal from the sink, each
node detects channel gain and transmits its signal immediately, if
its channel gain is above the threshold and the transmission is not
done in previous slots. In comparison, a node in deep fading (when
its channel gain is below the threshold) defers the transmission decision
to next slot, expecting a high channel gain in the future. If the
channel gain of a node is below the threshold over all slots, the
node transmits its signal in the last slot, and the signal arriving
at the sink is susceptible to magnitude misalignment.

To ensure a successful transmission, it is required that the channel
gain should be above the threshold $g_{th}$, over at least one of
the $N$ slots, with a relatively high probability $P_{th}$. On the
other hand, if channel gain of node $k$ is below $g_{th}$ over all
$N$ slots, node $k$ has to transmit its signal in the $N$-th slot,
with a probability $F_{g_{k}}(g_{th})^{N}$, where $F_{g_{k}}(x)$
denotes the CDF of channel gain $g_{k}$ per slot. Then, 

\begin{eqnarray}
\frac{1}{K}\sum_{k=1}^{K}P_{k}^{s}(g_{th}) & \ge & P_{th},\label{eq:cond-pdr}\\
P_{k}^{s}(g_{th}) & = & 1-F_{g_{k}}(g_{th})^{N},\nonumber 
\end{eqnarray}
should be satisfied. 

\begin{figure}
\centering

\includegraphics[width=6cm]{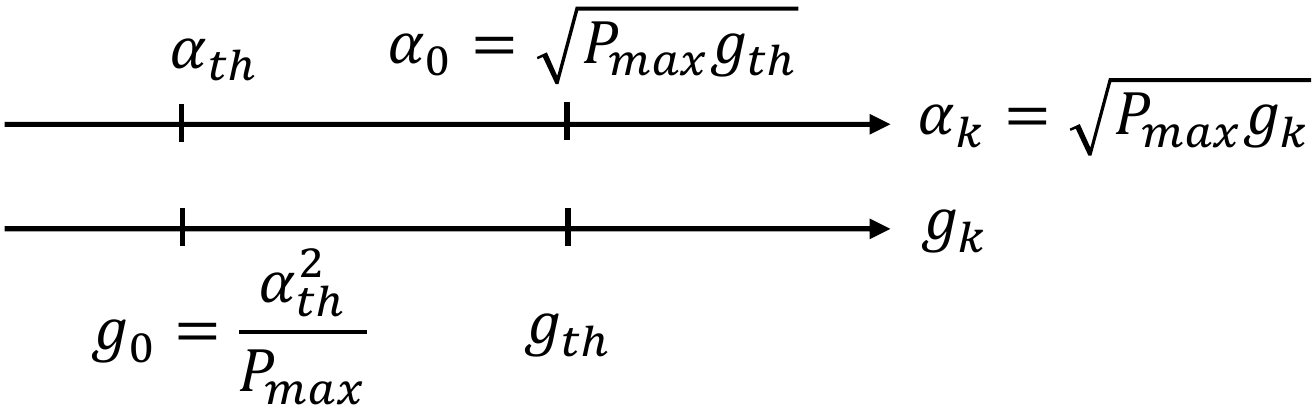}

\caption{\label{fig:analysis_alpha_g}Relationship between $\alpha_{th}$ ($g_{0}$)
and $\alpha_{0}$ ($g_{th}$).}
\end{figure}

Generally, at a given gain threshold $g_{th}$, using the maximal
power, the signal magnitude $\alpha_{0}=\sqrt{P_{max}g_{th}}$ should
be no less than $\alpha_{th}$. Figure~\ref{fig:analysis_alpha_g}
shows this relationship. Then, in the range $(\alpha_{0},\infty)$,
MSE of the signal part is

\begin{equation}
MSE_{1}=\sum_{k=1}^{K}\left(a\cdot\alpha_{th}\negmedspace-\negmedspace1\right)^{2}\negmedspace P_{k}^{s}(g_{th}).
\end{equation}
With a small probability $1-P_{k}^{s}(g_{th})$, the channel gain
of node $k$ is below the threshold $g_{th}$ over all slots, and
has to transmit its signal at the last slot, with a channel gain below
the threshold. In the range $(0,\alpha_{0})$, $\alpha_{k}$ has a
conditional distribution

\begin{equation}
f''_{\alpha_{k}}(x)=\frac{f_{\alpha_{k}}(x)}{F_{\alpha_{k}}(\alpha_{0})},\alpha_{k}<\alpha_{0}.
\end{equation}
This range is further divided into two sub-ranges. In the sub-range
$(\alpha_{th},\alpha_{0})$, the error term is $\left(a\cdot\alpha_{th}\negmedspace-\negmedspace1\right)^{2}$
with a conditional probability $\left(F_{\alpha_{k}}(\alpha_{0})\negmedspace-\negmedspace F_{\alpha_{k}}(\alpha_{th})\right)/F_{\alpha_{k}}(\alpha_{0})$.
In the sub-range $(0,\alpha_{th})$, the error term is $\left(a\cdot x\negmedspace-\negmedspace1\right)^{2}$.
Then, MSE of the signal part in the range $(0,\alpha_{0})$ is

\begin{eqnarray}
MSE_{2} & = & \sum_{k=1}^{K}MSE(k)\left(1-P_{k}^{s}(g_{th})\right),
\end{eqnarray}

\begin{align*}
MSE(k) & =\left(a\cdot\alpha_{th}\negmedspace-\negmedspace1\right)^{2}\negmedspace\frac{F_{\alpha_{k}}(\alpha_{0})\negmedspace-\negmedspace F_{\alpha_{k}}(\alpha_{th})}{F_{\alpha_{k}}(\alpha_{0})},\\
 & +\negmedspace\intop_{0}^{\alpha_{th}}\negmedspace\negmedspace\left(a\cdot x\negmedspace-\negmedspace1\right)^{2}f''_{\alpha_{k}}(x)dx.
\end{align*}
And the overall computation MSE in the whole range is

\begin{equation}
MSE=MSE_{1}+MSE_{2}+Na^{2}\sigma^{2}.\label{eq:mse-selfst}
\end{equation}
This is also a quadratic function of $a$, and the optimal $a$ for
the computation MSE can be computed by Eq.(\ref{eq:opt_a_selfst}).
Then, the optimal $\alpha_{th}$ is found via a grid search.

\begin{figure*}
\begin{equation}
a_{opt}=\frac{\sum_{k=1}^{K}\left(\alpha_{th}P_{k}^{s}(g_{th})+\alpha_{th}\frac{F_{\alpha_{k}}(\alpha_{0})\negmedspace-\negmedspace F_{\alpha_{k}}(\alpha_{th})}{F_{\alpha_{k}}(\alpha_{0})}\left(1-P_{k}^{s}(g_{th})\right)+\intop_{0}^{\alpha_{th}}xf''_{\alpha_{k}}(x)dx\cdot\left(1-P_{k}^{s}(g_{th})\right)\right)}{\sum_{k=1}^{K}\left(\alpha_{th}^{2}P_{k}^{s}(g_{th})+\alpha_{th}^{2}\frac{F_{\alpha_{k}}(\alpha_{0})\negmedspace-\negmedspace F_{\alpha_{k}}(\alpha_{th})}{F_{\alpha_{k}}(\alpha_{0})}\left(1-P_{k}^{s}(g_{th})\right)+\intop_{0}^{\alpha_{th}}x^{2}f''_{\alpha_{k}}(x)dx\cdot\left(1-P_{k}^{s}(g_{th})\right)\right)+N\sigma^{2}}.\label{eq:opt_a_selfst}
\end{equation}
\end{figure*}

Under Rayleigh fading, $g_{k}$, channel gain of node $k$, follows
an exponential distribution $f_{g_{k}}(x)$ with an average $\overline{g}_{k}$,
and has a CDF $F_{g_{k}}(x)$, as follows: 

\begin{equation}
f_{g_{k}}(x)=\frac{1}{\overline{g}_{k}}\exp\left(-\frac{x}{\overline{g}_{k}}\right),
\end{equation}

\begin{equation}
F_{g_{k}}(x)=1-\exp\left(-\frac{x}{\overline{g}_{k}}\right).
\end{equation}
Then, $\alpha_{k}=\sqrt{P_{max}g_{k}}$ has a PDF $f_{\alpha_{k}}(x)$
and a CDF $F_{\alpha_{k}}(x)$,

\begin{equation}
f_{\alpha_{k}}(x)=\frac{2x}{P_{max}\overline{g}_{k}}\exp\left(-\frac{x^{2}}{P_{max}\overline{g}_{k}}\right),
\end{equation}

\begin{equation}
F_{\alpha_{k}}(x)=1-\exp\left(-\frac{x^{2}}{P_{max}\overline{g}_{k}}\right).
\end{equation}

\subsubsection{Relation between $a$ and $\alpha_{th}$ \label{sec:Relation-a-alphath}}

Although $\alpha_{th}$ and $a$ are separately computed, they are
highly correlated. Actually, their product, $a\cdot\alpha_{th}$,
is approximately 1.0. 

Now look back at Eq.(\ref{eq:mse-optsel}). It is straightforward
that $a\cdot\alpha_{th}$ approaching 1.0 will ensure that signals
aligned to $\alpha_{th}$ have a computation error close to 0. In
fact, it is assumed $a\cdot\alpha_{th}=1.0$ in \cite{Liu20}. But
in the analysis, assuming $a\cdot\alpha_{th}=1$ will make the whole
process fail. Once the optimal parameters are determined, it is safe
to use the approximation $a\cdot\alpha_{th}=1$ to compute the actual
computation MSE.

\subsubsection{The whole process of the proposed method}

It is summarized as follows: 
\begin{itemize}
\item For each possible number of slots, $N$, $g_{th}$ is computed from
$P_{th}$ according to Eq.(\ref{eq:cond-pdr}), which can be simplified
in Rayleigh fading.
\item For each possible $\alpha_{th}$, the optimal $a$ is computed via
Eq.(\ref{eq:opt_a_selfst}). 
\item Parameters ($N,g_{th},\alpha_{th}$) leading to the minimal computation
MSE are found.
\item The sink node broadcasts a beacon signal (containing $N,g_{th},\alpha_{th}$)
at the beginning of each slot. 
\item Each node monitors the beacon signal, obtains its channel gain $g_{k}$,
and compares it with the threshold $g_{th}$. If $g_{k}\ge g_{th}$
and the node has not transmitted its signal yet, it transmits its
signal immediately. Otherwise, if the channel gain remains below the
threshold over all $N$ slots, a node transmits its signal in the
$N$-th slot. The transmission power of node $k$ is computed as $b_{k}=min\{\alpha_{th}/h_{k},\sqrt{P_{max}}\}$,
using the channel coefficient $h_{k}$. 
\item At the end of the $N$-th slot, the signals received over all $N$-slots
are summed together at the sink.
\end{itemize}

\subsection{Selecting the optimal slot}

Under the generated case, channel gains of all slots are known in
advance and each node will select the slot with the maximal gain to
transmit its signal. This method determines the performance upper
bound of the proposed method. 

According to order statistics \cite{David03}, $f_{\alpha_{k}}^{N}(x)$
is 

\begin{equation}
f_{max(\alpha_{k})}(x)=N\cdot f_{\alpha_{k}}(x)\cdot F_{\alpha_{k}}(x)^{N-1}.
\end{equation}
Then, the optimal parameters minimizing the computation MSE in Eq.(\ref{eq:mse-optsel})
can be found.

\subsection{Further analysis}

Here, we further analyze the average channel gain, the probability
of misalignment in signal magnitude, and transmission power.

\subsubsection{Average channel gain}

Average channel gain in the proposed method is

\begin{align}
\frac{1}{K}\sum_{k=1}^{K}P_{k}^{s}(g_{th})\negmedspace\cdot\negmedspace\int_{g_{th}}^{\infty}\negmedspace\negmedspace x\cdot f'_{g_{k}}(x)dx\nonumber \\
+\left(1\negmedspace-\negmedspace P_{k}^{s}(g_{th})\right)\negmedspace\int_{0}^{g_{th}}\negmedspace\negmedspace x\cdot f''_{g_{k}}(x)dx & ,
\end{align}

\[
f'_{g_{k}}(x)=\frac{f_{g_{k}}(x)}{1-F_{g_{k}}(g_{th})},g_{k}\ge g_{th},
\]

\[
f''_{g_{k}}(x)=\frac{f_{g_{k}}(x)}{F_{g_{k}}(g_{th})},g_{k}<g_{th}.
\]
Average channel gain in the optimal selection is 

\begin{align}
 & \frac{1}{K}\sum_{k=1}^{K}\int_{0}^{\infty}x\cdot f_{max(g_{k})}(x)dx,\\
 & f_{max(g_{k})}(x)=N\cdot f_{g_{k}}(x)\cdot F_{g_{k}}(x)^{N-1}(x).\nonumber 
\end{align}

\subsubsection{Probability of misalignment in signal magnitude\label{sec:Prob-misalign}}

In the proposed method, let $g_{0}$ denote the channel gain $\frac{\alpha_{th}^{2}}{P_{max}}$.
The signal of node $k$ arrives at the sink with signal magnitude
in alignment when $g_{k}\ge g_{0}$. Considering the two ranges $(g_{th},\infty)$
and $(0,g_{th})$, the overall probability that the signal from node
$k$ is in alignment is

\[
P_{k}^{s}(g_{0})=P_{k}^{s}(g_{th})+\left(1-P_{k}^{s}(g_{th})\right)\frac{F_{g_{k}}(g_{th})\negmedspace-\negmedspace F_{g_{k}}(g_{0})}{F_{g_{k}}(g_{th})}.
\]
And the probability that the signal magnitude misalignment occurs
in one or more signals is

\begin{equation}
1-\prod_{k=1}^{K}P_{k}^{s}(g_{0}).
\end{equation}
When $p=1-P_{k}^{s}(g_{0})$ is the same for all nodes, the probability
that signals of $k$ nodes are misaligned in signal magnitude is

\begin{equation}
p_{k}=\left(\begin{array}{c}
K\\
k
\end{array}\right)p^{k}(1-p)^{K-k}.
\end{equation}
Then, the average number of signals with magnitude misalignment is
$\sum_{k}k\cdot p_{k}$.

In the optimal selection method, the probability that the signal of
node $k$ arrives at the sink with signal magnitude in alignment is
$1-F_{\alpha_{k}}^{N}(\alpha_{th})$. Then, the probability that the
signal magnitude misalignment occurs in one or more signals is 

\begin{equation}
1-\prod_{k=1}^{K}\left(1-F_{\alpha_{k}}^{N}(\alpha_{th})\right).
\end{equation}
When $p=F_{\alpha_{k}}^{N}(\alpha_{th})$ is the same for all nodes,
the probability that signals of $k$ nodes are misaligned in signal
magnitude can be computed in a similar way as in the proposed method.

\subsubsection{Average transmission power}

In the proposed method, the power required for transmission depends
on the channel gain threshold. For a node $k$ with a gain above the
threshold ($g_{k}\ge g_{th}$), the average transmission power of
node $k$, $b_{k}^{2}=\alpha_{th}^{2}/h_{k}^{2}=\alpha_{th}^{2}/g_{k}$,
is computed as 

\begin{equation}
E_{k1}=\int_{g_{th}}^{\infty}\frac{\alpha_{th}^{2}}{g_{k}}\cdot f'{}_{g_{k}}(x)dx.
\end{equation}

Next consider a node $k$ whose channel gain is below the threshold
$g_{th}$. When $g_{k}\in(g_{0},g_{th})$, $b_{k}=\alpha_{th}/\sqrt{g_{k}}$
is less than $\sqrt{P_{max}}$, and the channel inversion policy is
applied for transmission power control. On the other hand, when $g_{k}\in(0,g_{0}]$,
the maximal power $P_{max}$ is used. On average, the transmission
power is 

\begin{equation}
E_{k2}=\int_{0}^{g_{0}}P_{max}\cdot f''_{g_{k}}(x)dx+\int_{g_{0}}^{g_{th}}\frac{\alpha_{th}^{2}}{g_{k}}\cdot f''_{g_{k}}(x)dx.
\end{equation}
Then, the average transmission power of all nodes can be computed
by 

\begin{equation}
\frac{1}{K}\sum_{k=1}^{K}P_{k}^{s}(g_{th})\cdot E_{k1}+\left(1-P_{k}^{s}(g_{th})\right)\cdot E_{k2}.
\end{equation}

In the optimal selection method, the average transmission power is
equal to

\begin{equation}
\frac{1}{K}\sum_{k=1}^{K}\int_{\alpha_{th}}^{\infty}\frac{\alpha_{th}^{2}}{x}\cdot f_{max(g_{k})}(x)dx+\int_{0}^{\alpha_{th}}P_{max}\cdot f_{max(g_{k})}(x)dx.
\end{equation}

\subsection{Numerical Analysis}

Here, we do some numerical analysis, using the default setting in
Table~\ref{tab:SimCond} unless specified otherwise. Here, all links
have the same average channel gain ($g_{k}=10$) and instantaneous
channel gain follows independent fast Rayleigh fading. In the analysis,
the optimal selection method is denoted as OptSel, and the proposed
method is denoted as SelFirst.

\begin{figure}
\centering

\includegraphics[width=8cm]{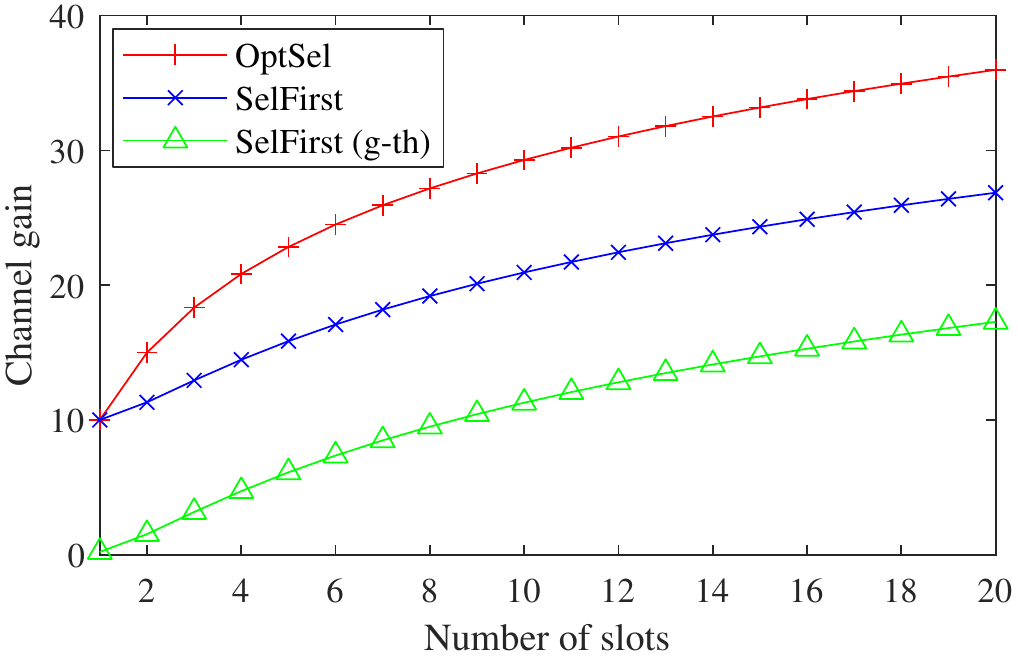}

\caption{\label{fig:Ana-chgain}Channel gain and its threshold under different
numbers of slots ($\overline{g}_{k}=10$dB, $P_{th}=0.98$, $P_{max}=10$).}
\end{figure}

Figure~\ref{fig:Ana-chgain} shows the average channel gain and the
channel gain threshold (SelFirst(g-th)) under different numbers of
slots. At $N$=1 slot, the average channel gain is 10, in both SelFirst
and OptSel. Then, channel gain increases with the number of slots,
although there is a gap between SelFirst and OptSel. The increase
of channel gain in SelFirst is because the threshold of channel gain
increases with the number of slots. The gap between SelFirst and OptSel
is because SelFirst does not know the channel gain of all slots in
advance, and setting the threshold cannot ensure to select the optimal
slot with the maximal gain, although it does help avoid deep fading. 

\begin{figure}
\centering

\includegraphics[width=8cm]{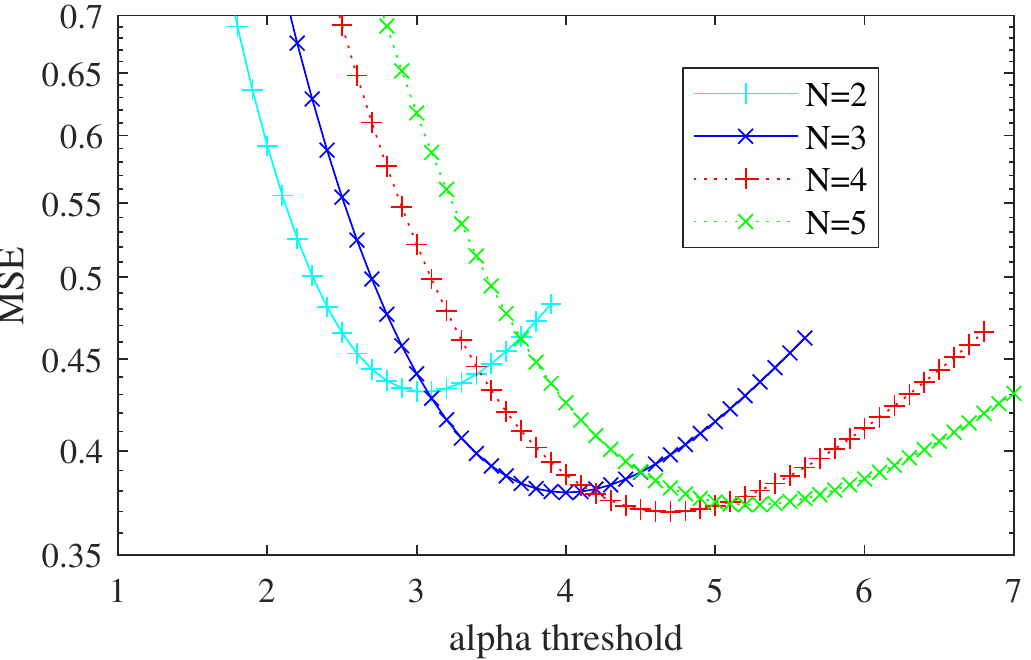}

\caption{\label{fig:Ana-mse-alphath}Variation of computation MSE with respect
to $\alpha_{th}$ ($\overline{g}_{k}=10$dB, $P_{th}=0.98$, $P_{max}=10$). }
\end{figure}

Figure~\ref{fig:Ana-mse-alphath} shows how the computation MSE varies
with $\alpha_{th}$ in the SelFirst method. Clearly, the computation
MSE reaches a minimum at some place, which depends on $N$, the number
of slots.

\begin{figure}
\centering

\includegraphics[width=8cm]{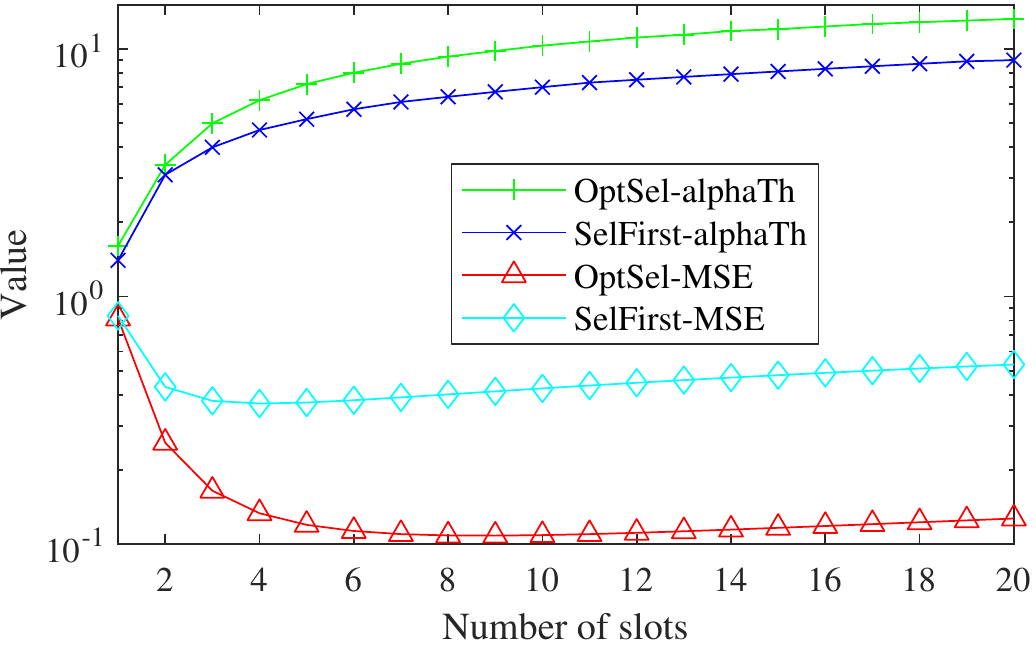}

\caption{\label{fig:Ana-alpha-mse}Change of $\alpha_{th}$ and computation
MSE with respect to $N$, the number of slots ($\overline{g}_{k}=10$dB,
$P_{th}=0.98$, $P_{max}=10$).}
\end{figure}

We further investigate the impact of $N$, the number of slots. As
shown in Figure~\ref{fig:Ana-alpha-mse}, with the increase of $N$,
$\alpha_{th}$ increases in both OptSel and SelFirst. Meanwhile the
computation MSE first decreases, and after reaching the minimum, increases
again, which indicates that some $N$ is optimal. OpSel has a larger
$\alpha_{th}$, and accordingly a smaller computation MSE.

\begin{figure}
\centering

\includegraphics[width=8cm]{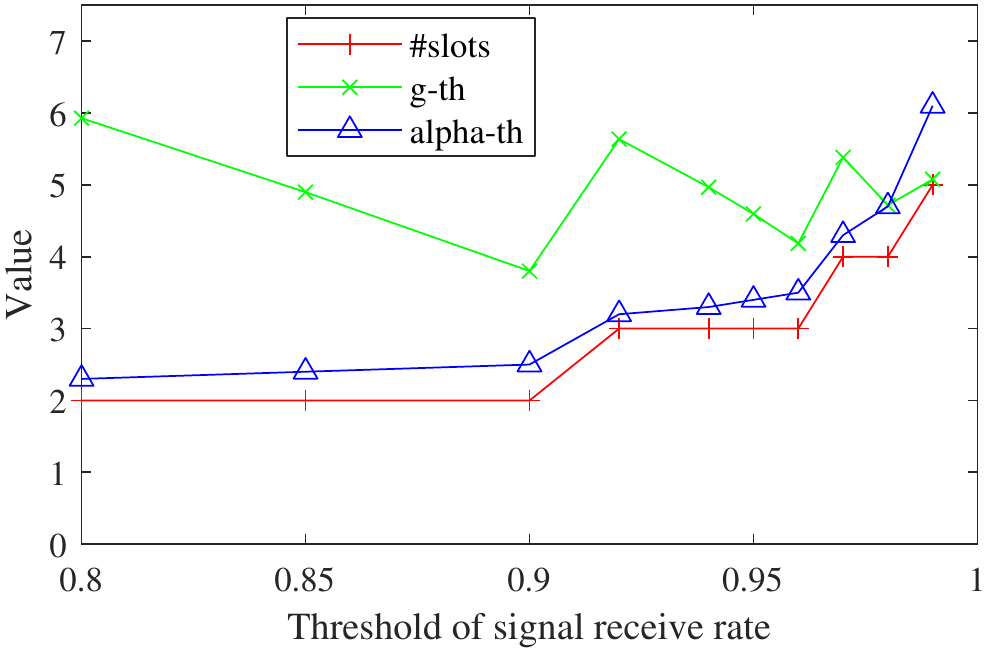}

\caption{\label{fig:Ana-opt-param}Optimal parameters: $N$, $g_{th}$ and
$\alpha_{th}$ under different $P_{th}$ ($\overline{g}_{k}=10$dB,
$P_{max}=10$).}
\end{figure}

Figure~\ref{fig:Ana-opt-param} shows how optimal parameters ($N,$
$g_{th}$, $\alpha_{th}$) change with $P_{th}$. With the increase
of $P_{th}$, the number of slots gradually increases. At the same
number of slots, $g_{th}$ decreases when $P_{th}$ increases, because
a smaller $g_{th}$ ensures more nodes transmit their signals to meet
the requirement of $P_{th}$. $\alpha_{th}$ always increases with
$P_{th}$, which helps to reduce $a$ in the MSE equation, and reduce
the impact of noise, but at the cost of more transmission power because
transmission power is proportional to $\alpha_{th}^{2}$.

The optimal parameters are shown in Table~\ref{tab:OptParam}.

\begin{table}[!t]
\caption{Optimal parameters under different $P_{th}$.}

\label{tab:OptParam} 

\begin{centering}
\begin{tabular}{l|cccccccc}
\hline 
$P_{th}$ & 0.90 & 0.92 & 0.94 & 0.95 & 0.96 & 0.97 & 0.98 & 0.99\tabularnewline
\hline 
$N$ & 2 & 3 & 3 & 3 & 3 & 4 & 4 & 5\tabularnewline
$\alpha_{th}$ & 2.5 & 3.2 & 3.3 & 3.4 & 3.5 & 4.3 & 4.7 & 6.1\tabularnewline
$g_{th}$ & 3.8 & 5.6 & 5.0 & 4.6 & 4.2 & 5.4 & 4.7 & 5.1\tabularnewline
\hline 
\end{tabular}
\par\end{centering}
\end{table}

\section{Simulation Evaluation\label{sec:Evaluation}}

Here, we evaluate the proposed SelFirst method by Monte Carlo simulation,
and compare it with the OptSel method, and the AirComp method \cite{Liu20}
described in Sec~\ref{sec:Basic-AirComp}. The number of run is 100,000
for each setting.

We mainly consider three metrics, transmission power, the computation
MSE, and the probability of misalignment in signal magnitude. It is
assumed that the instantaneous channel gains are known to the sink
in AirComp. In comparison, the instantaneous channel gains are not
known to the sink in the proposed SelFirst method. To make a fair-comparison,
OptSel uses the same number of slots as SelFirst does.

First, we use a scenario where all channels follow Rayleigh fading
and have the same average channel gain $\overline{g}_{k}=10$. The
parameters in Table~\ref{tab:OptParam} are used for SelFirst.

\begin{figure}
\centering

\includegraphics[width=8cm]{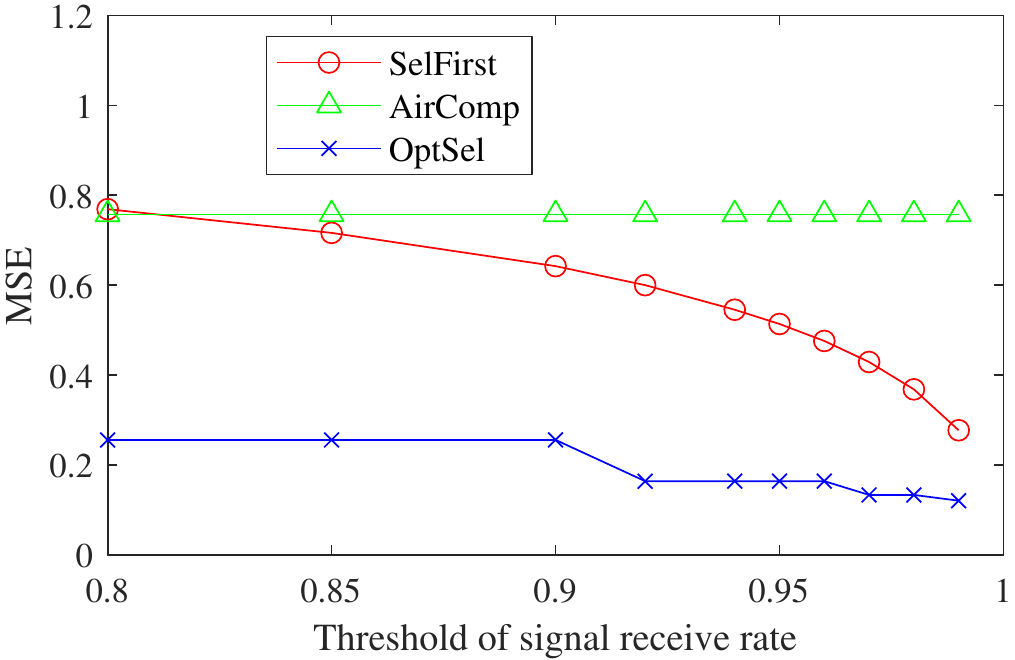}

\caption{\label{fig:Res-SSR-mse}Computation MSE under different $P_{th}$
($\overline{g}_{k}=10$dB, $P_{max}=10$).}
\end{figure}

Figure~\ref{fig:Res-SSR-mse} shows the computation MSE in three
methods, under different $P_{th}$. Generally, SelFirst lies between
OptSel and AirComp. At $P_{th}=0.8$, SelFirst and AirComp have almost
the same computation MSE. But as $P_{th}$ increases, the computation
MSE in the SelFirst method gradually decreases and approaches that
of OptSel. 

\begin{figure}
\centering

\includegraphics[width=8cm]{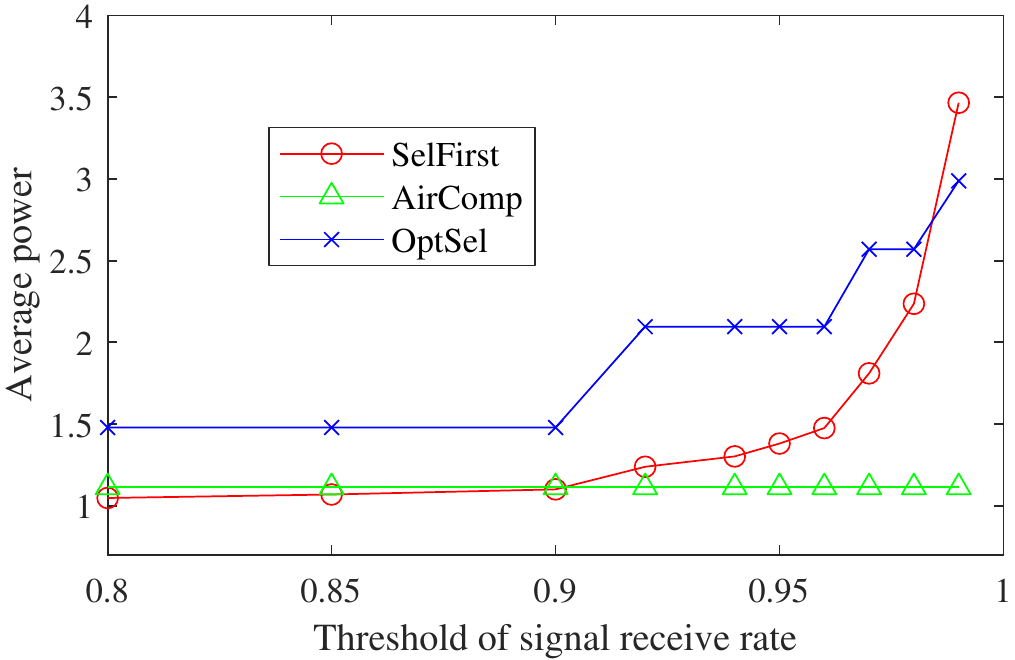}

\caption{\label{fig:Res-SSR-pow}Transmission power under different $P_{th}$
($\overline{g}_{k}=10$dB, $P_{max}=10$).}
\end{figure}

Figure~\ref{fig:Res-SSR-pow} shows the transmission power in three
methods. Here, the transmission power in SelFirst and OptSel increases
with $P_{th}$. This is consistent with the result of $\alpha_{th}$
in Figure~\ref{fig:Ana-opt-param}. Surprisingly, both SelFirst and
OptSel consume more power than AirComp. It is clear that a tradeoff
between the computation MSE and transmission power is necessary.

\begin{figure}
\centering

\includegraphics[width=8cm]{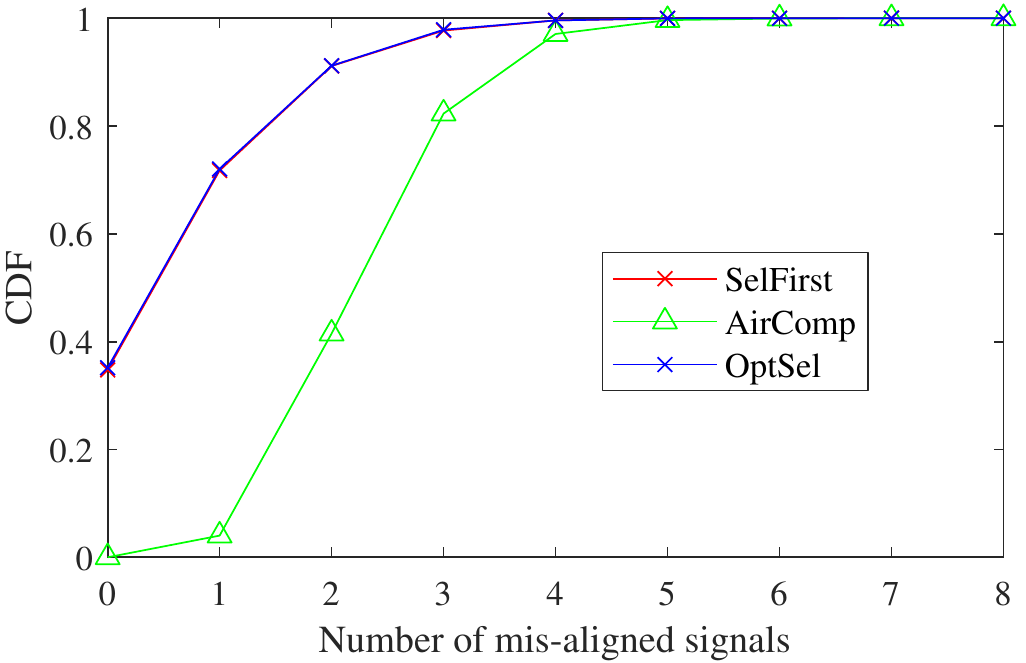}

\caption{\label{fig:Res-per-cdf}Cumulative probability of signal misalignment
up to $k$ signals ($\overline{g}_{k}=10$dB, $P_{th}=0.98$, $P_{max}=10$).}
\end{figure}

Next, we investigate the probability of misalignment in signal magnitude.
Figure~\ref{fig:Res-per-cdf} shows the cumulative probability of
signal misalignment up to $k$ signals, and the value $k$ corresponds
to the horizontal axis. Here, SelFirst and OptSel have almost the
same performance. The average number of signals with signal magnitude
misalignment is 1.05 in SelFirst, 1.04 in OptSel, while 2.75 in AirComp.
These results are consistent with the analysis in Sec~\ref{sec:Prob-misalign}.

\section{Tradeoff between transmission power and MSE\label{sec:tradeoff-pow-mse}}

Results in Figure~\ref{fig:Res-SSR-mse} and Figure~\ref{fig:Res-SSR-pow}
have shown that a tradeoff is necessary between the computation MSE
and transmission power. 

\begin{figure}
\centering

\includegraphics[width=8cm]{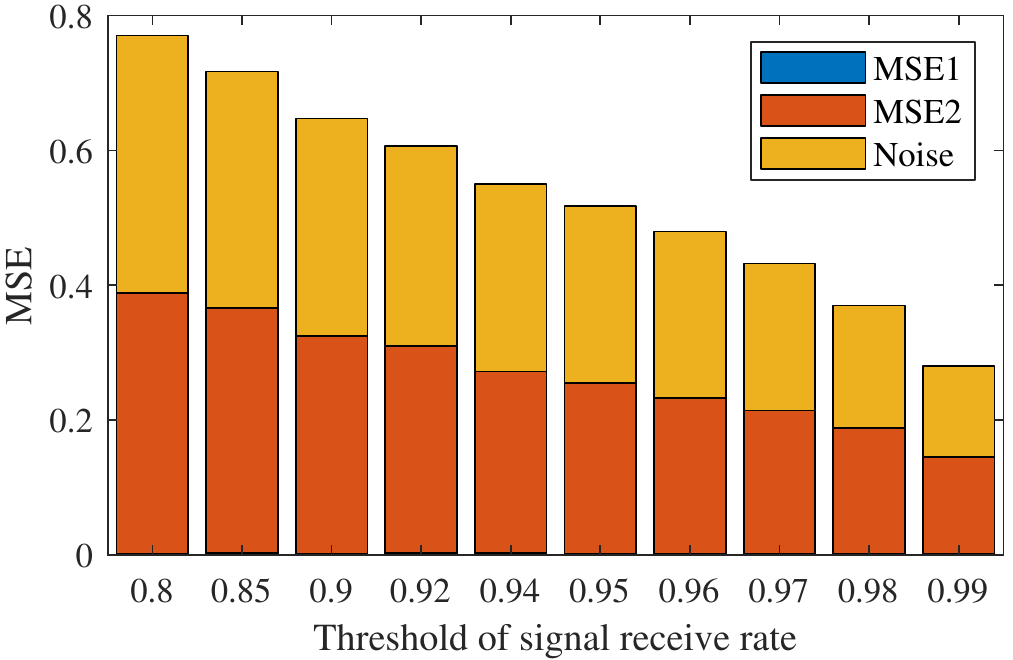}

(a) Before control

\includegraphics[width=8cm]{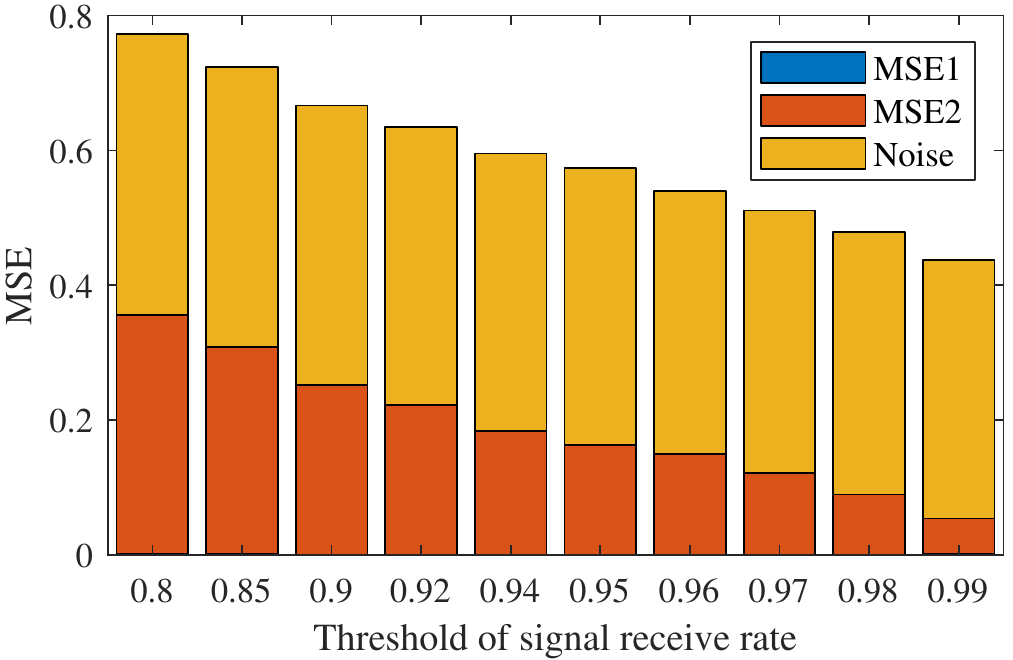}

(b) After control

\caption{\label{fig:Res-mse-part}Different components of computation MSE under
different $P_{th}$ ($\overline{g}_{k}=10$dB, $P_{max}=10$). MSE1
is very small and invisible in the figure.}
\end{figure}

In the proposed SelFirst method, we investigate the three parts of
MSE, MSE1 (signals with channel gain above the threshold), MSE2 (signals
with channel gain below the threshold) and noise in Eq.(\ref{eq:mse-selfst}).
Figure~\ref{fig:Res-mse-part}(a) shows the result. Generally, MSE1
and MSE2 are reduced by transmission power control. Here, MSE1 is
almost negligibly small. By setting a threshold of channel gain ($g_{th}$
increases with $P_{th}$), MSE2 is further reduced. But surprisingly,
noise is also reduced, although it has a factor of $N$($N$ increases
with $P_{th}$). 

To explain this, we look back at Eq.(\ref{eq:Nslot-mse}). MSE of
the signal part depends on $a\cdot\alpha_{th}$, and MSE of the noise
part is $Na^{2}\sigma^{2}$. Reducing the MSE of the noise part will
lead to a small $a$. Because $a\cdot\alpha_{th}$ approaches 1.0
(to reduce the computation error of the signal part, as discussion
in Sec.~\ref{sec:Relation-a-alphath}), this will lead to a large
$\alpha_{th}$ and accordingly a large transmission power (which is
proportional to $\alpha_{th}^{2}$). 

To solve this problem, we propose that the optimization is focused
on the signal part, and avoid the over-reduction of the noise part.
Specifically, $\max\{N\cdot a^{2},\beta\}\cdot\sigma^{2}$ is used
instead of $Na^{2}\sigma^{2}$ in Eq.(\ref{eq:mse-selfst}) so that
the optimization for the noise part will stop once the noise power
is reduced to $\beta\sigma^{2}$, where $\beta$ is an adjustable
parameter. It should be noted that this is only used for finding the
optimal parameters, not for the computation of MSE in the experiment.
By setting $\beta$ to 0.4 in SelFirst, the corresponding MSE parts
are shown in Figure~\ref{fig:Res-mse-part}(b). Here, the noise part
is almost fixed while MSE of the signal part is reduced more compared
with the result in Figure~\ref{fig:Res-mse-part}(a).

\begin{figure}
\centering

\includegraphics[width=8cm]{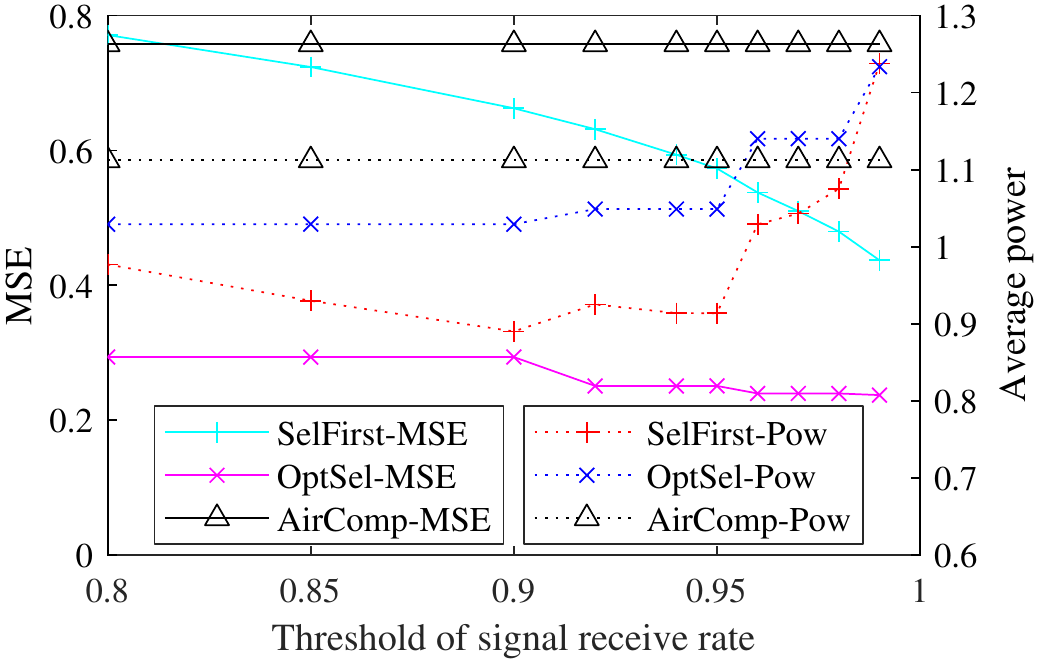}

\caption{\label{fig:Res-mse-pow-tr}Computation MSE and transmission power
under different $P_{th}$ ($\overline{g}_{k}=10$dB, $P_{max}=10$).}
\end{figure}

The overall computation MSE and transmission power of the three methods
are shown in Figure~\ref{fig:Res-mse-pow-tr}. Here, $\beta=0.25$
is used in OptSel. As $P_{th}$ increases, the reduction of the computation
MSE by SelFirst becomes smaller compared with the result in Figure~\ref{fig:Res-SSR-mse},
but the quick increase of transmission power is avoided, which confirms
that the simple method is effective in balancing transmission power
and the computation MSE.  At $P_{th}=0.98$, SelFirst consumes almost
the same transmission power as AirComp, but reduces the computation
MSE by 36.7\%.

\subsection{Evaluation in a scenario with non equal channel gains}

Next we evaluate the three methods in a general scenario where all
nodes have different average channel gains. Specifically, all nodes
are uniformly distributed in a square area (200m x 200m), and the
sink node is located at the center. The mean value of average channel
gains in dB is 10, and each node has a channel gain depending on its
location, which also varies with time according to Rayleigh fading.

\begin{figure}
\centering

\includegraphics[width=8cm]{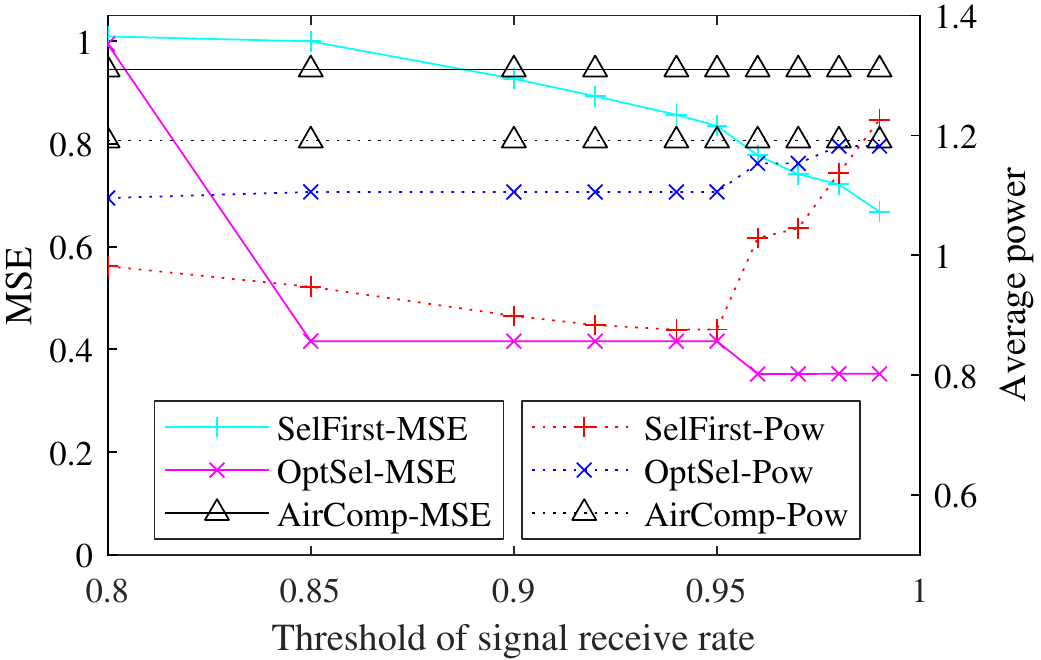}

\caption{\label{fig:Res-mse-pow-gen}Computation MSE and transmission power
under different $P_{th}$ in a general scenario where average channel
gain of a node varies with its location (The mean value of average
channel gains in dB is 10, $P_{max}=10$).}
\end{figure}

Figure \ref{fig:Res-mse-pow-gen} shows how transmission power and
compuation MSE change with $P_{th}$ in three methods. This has a
similar trend as in Figure \ref{fig:Res-mse-pow-tr}, but the variation
of average channel gains has a little negative impact on the performance.
At $P_{th}=0.98$, SelFirst still consumes almost the same transmission
power as AirComp, but reduces the computation MSE only by 23.6\%.
This result is still promising considering that the sink needs to
know instantaneous channel gains of all nodes in AirComp but only
need to know channel statistics in SelFirst.

\section{Conclusion \label{sec:Conclusion}}

To deal with fast channel fading in over-the-air computation, this
paper has extended the conventional AirComp method, by distributing
data transmission/fusion to multiple slots. A threshold of channel
gain is set to make a node transmit its signal only when its channel
gain gets above the threshold in any slot, or in the last slot if
its channel gain remains below the threshold. This helps to avoid
deep fading and improve channel gains, which further facilitate the
transmission power control for the alignment of signal magnitudes.
Theoretical analysis gives the closed-form of the computation MSE,
which helps to find optimal parameters. The transmission scheduling
at each node is conducted in a distributed way, and works well in
fast fading environment. By avoiding over-reducing MSE of the noise
part, the proposed method reduces the computation MSE while consuming
less or almost the same power as AirComp.

The same threshold of channel gain is used, even though nodes have
different average channel gains, in order to ensure the alignment
of signal magnitude. Then, signals of nodes far away from the sink
have a larger chance of signal magnitude misalignment than those of
nodes near the sink. Relay is an effective method for this case, and
will be studied in the future work. 

\bibliographystyle{IEEEtran}
\bibliography{IEEEabrv,mybibfile}

\end{document}